\documentclass[twocolumn,french,english,english,pra,american,amsmath,amssymb,showpacs,superscriptaddress]{revtex4}

\pdfoutput = 1

\usepackage[T1]{fontenc}
\usepackage[latin9]{inputenc}
\usepackage{color}
\usepackage{amsmath}
\usepackage{amssymb}
\usepackage{graphicx}
\usepackage{esint}

\makeatletter
\@ifundefined{textcolor}{}
{%
 \definecolor{BLACK}{gray}{0}
 \definecolor{WHITE}{gray}{1}
 \definecolor{RED}{rgb}{1,0,0}
 \definecolor{GREEN}{rgb}{0,1,0}
 \definecolor{BLUE}{rgb}{0,0,1}
 \definecolor{CYAN}{cmyk}{1,0,0,0}
 \definecolor{MAGENTA}{cmyk}{0,1,0,0}
 \definecolor{YELLOW}{cmyk}{0,0,1,0}
}

\@ifundefined{definecolor}
 {\usepackage{color}}{}
\newcommand{\noshow}[1]{}

\makeatother
\usepackage[colorlinks,bookmarks=false,citecolor=blue,linkcolor=red,urlcolor=blue]{hyperref}

\makeatother

\usepackage{babel}
\makeatletter
\addto\extrasfrench{%
   \providecommand{\fg}{\ifdim\lastskip>\z@\unskip\fi~\frqq}%
}

\makeatother
\begin{document}

\preprint{This line only printed with preprint option}

\title{Emergence of nonlinear behavior in the dynamics of ultracold bosons}

\author{Beno\^{i}t Vermersch}

\altaffiliation{Present Address: Institute for Quantum Optics and Quantum Information of the Austrian Academy of Sciences, A-6020 Innsbruck, Austria}

\selectlanguage{french}%

\affiliation{Laboratoire de Physique des Lasers, Atomes et Mol{\'e}cules, Universit\foreignlanguage{english}{{\'e}}
Lille 1 Sciences et Technologies, CNRS; F-59655 Villeneuve d'Ascq
Cedex, France}

\selectlanguage{english}%

\author{Jean~Claude Garreau}

\selectlanguage{french}%

\affiliation{Laboratoire de Physique des Lasers, Atomes et Mol{\'e}cules, Universit\foreignlanguage{english}{{\'e}}
Lille 1 Sciences et Technologies, CNRS; F-59655 Villeneuve d'Ascq
Cedex, France}

\homepage{http://www.phlam.univ-lille1.fr/atfr/cq}

\begin{abstract}
We study the evolution of a system of interacting ultracold bosons,
which presents nonlinear, chaotic, behaviors in the limit of very
large number of particles. Using the spectral entropy as an indicator
of chaos and three different numerical approaches: Exact diagonalization,
truncated Husimi method and mean-field (Gross-Pitaevskii) approximation,
we put into evidence the destructive impact of quantum noise on the
emergence of the nonlinear dynamics.
\end{abstract}
\maketitle

\section{Introduction}

Ultracold atoms are clean, controllable, highly flexible experimental
systems, that can often be modeled from first principles. For such
reasons they have become in recent years a preferred testing ground
for quantum many-body effects~\cite{Bloch:ManyBodyUltracold:RMP08}.
A particularly interesting example is the emergence of chaotic behaviors
in quantum systems: The Schr{\"o}dinger equation is linear and hence
cannot display chaotic behavior in the classical sense (i.e. chaos
associated to a sensitivity to initial conditions), however, the classical
world very often display chaos. Bose-Einstein condensates can be produced
in laboratories at mesoscopic sizes (up to $10^{8}$ atoms)~\cite{Streed:LargeAtomNumberBEC:RSI06}
in intrinsically quantum-coherent states, they are thus ideal systems
for the study of the transition form quantum to classical behavior,
and, in particular, the emergence of chaotic behaviors. The corresponding
quantum many-body problem (with binary contact interactions) can be
treated by decomposing the matter-wave field $\hat{\psi}$ in a macroscopic
part describing a single particle wave function $\psi(x,t)$, called
``condensed fraction'', plus a fluctuating quantum field corresponding
to the excitations of the matter-wave field. The wave function $\psi(x,t)$
obeys the well-known Gross-Pitaevskii equation (GPE), which includes
a \emph{nonlinear} term~\cite{Stringari:BECRevTh:RMP99,Cohen-TannoudjiDGO:AdvancesInAtomicPhysics::11,PethickSmith:BoseEinstein:08}:
\begin{equation}
i\hbar\frac{\partial\psi(x,t)}{\partial t}=\left(H_{1}+gN\left|\psi(x,t)\right|^{2}\right)\psi(x,t),\label{eq:GPE}
\end{equation}
where $g$ is the nonlinear parameter proportional to the $s$-wave
scattering length,\textcolor{blue}{{} }$N$ is the number of particles
and $H_{1}$ is the one-particle Hamiltonian. The nonlinear term models
the ``mean effect'' of particle-particle interactions, and is simply
proportional to the spatial probability density $\left|\psi(x,t)\right|^{2}$~%
\footnote{The form~(\ref{eq:GPE}) of the GPE corresponds to the normalization
choice $\int dx\left|\psi\right|^{2}=1$.%
}. Replacing the many-body operator $\hat{\psi}$ by a single-particle
wave function $\psi$ which leads to the effective equation~(\ref{eq:GPE})
implies neglecting quantum fluctuation, but preserves the long-range
order supposed to be an essential property of condensates. It turns
out that in a large variety of interesting situations this rather
rough approach gives a very good description of the condensate dynamics
\cite{Stringari:BECRevTh:RMP99}. It can nevertheless be surprising
that a linear exact problem can be accurately modeled by a nonlinear
effective equation which may present a qualitatively different dynamics
(e.g. sensitivity to initial conditions and chaos). The mean-field
approximation hides the ``microscopic'' origin of the nonlinearity,
and one is left with a situation similar to that encountered in statistical
physics: The macroscopic dynamics of a gas can present qualitatively
different characteristics (e.g. irreversibility) from the microscopic
dynamics of individual molecules (which is reversible). The averaging
over the microscopic dynamics leads to a loss of symmetry and to a
qualitatively different behavior in the macroscopic scale. One can
speculate that the nonlinear behavior of a condensate arises in an
analog way: Averaging over the microscopic (quantum) dynamics produces
a qualitatively different behavior. A particularly spectacular manifestation
of this is the ``quasiclassical'' chaos that can appear in the dynamics
of a condensate~\cite{Thommen:ChaosBEC:PRL03,Smerzi:InstabilityBEC:PRL04,Lepers:QuasiClassTrack:PRL08,Fallani:InstabilityBEC:PRL04,Trimborn:MeanFieldlBoseHubbard:PRA09},
that is, chaos related to sensitivity to initial conditions. In the
present work, we will use a simple model displaying quasiclassical
chaos in the mean-field approximation and will compare it to the solutions
of the many-body problem. By adequately choosing the representation
of the dynamics we will show that one can observe ``traces'' of
the quasiclassical chaos in the behavior of the many-body system even
with a limited number of atoms.

\section{The model}

The toy model used here consists in a ultracold boson gas placed in
an accelerated (or tilted) optical potential~\cite{BenDahan:BlochOsc:PRL96,Raizen:WSOptPot:PRL96}
corresponding to the single-particle Hamiltonian
\begin{equation}
H_{1}=-\frac{1}{\pi^{2}}\frac{\partial^{2}}{\partial x^{2}}+V_{0}\cos(2\pi x)+Fx\label{eq:H1WS}
\end{equation}
where we used normalized units~\cite{Thommen:ChaosBEC:PRL03} such
that the lattice constant of the periodic part of the potential is
1, energies are measured in units of the so-called recoil energy ($\hbar\omega_{r}=\hbar^{2}k_{L}^{2}/2M$,
with $k_{L}=\pi/d$ where $d$ is the lattice constant of the periodic
part of the potential and $M$ is the mass of the atoms) and we have
set $\hbar=1$. In presence of interactions, a nonlinear term $g\left|\psi(x,t)\right|^{2}$
is added to the above Hamiltonian\textcolor{red}{{} }leading to Eq.
\eqref{eq:GPE}. We confine the dynamics to three adjacent sites $s=-1,0,1$
placed at $x=-1,0,1$ with Dirichlet boundary conditions~%
\footnote{This is not an unphysical assumption, for if $g$ is not too small,
the nonlinear term decouples the dynamics of populated wells from
the others (a phenomenon called \emph{self-trapping}) and the population
can stay confined in these three wells for quite long times.%
} and restrict the dynamics to the lowest band of the system; one can
then write the solution of Eq.~(\ref{eq:H1WS}) in the form
\begin{equation}
\psi(x,t)=c_{-1}e^{iFt}w(x+1)+c_{0}w(x)+c_{1}e^{-iFt}w(x-1)\label{eq:WSsolution}
\end{equation}
where $w(x)$ is the Wannier-Stark function {[}eigenstate of~(\ref{eq:H1WS}){]}
of the fundamental ladder %
\footnote{As long as $gN\int w^{4}(x)dx\ll V_{0}$ the coupling to higher ladders
is negligible.%
} associated to the $x=0$ site~\cite{Thommen:WannierStark:PRA02}.
In the case $g\neq0$\textcolor{blue}{{} }, we keep the above form,
but we let the coefficients $c_{s}$ depend on time. Then, inserting
Eq.~(\ref{eq:WSsolution}) in Eq.~(\ref{eq:GPE}) one obtains a
set of coupled differential equations that can be integrated numerically:
\begin{eqnarray}
i\dot{c_{s}} & = & Fsc_{s}+UN|c_{s}|^{2}c_{s}\nonumber \\
 & + & J_{\pm}N\left(2|c_{s}|^{2}c_{s\pm1}+|c_{s\mp1}|^{2}c_{s\mp1}+c_{s\pm1}^{*}c_{s}^{2}\right)\label{eq:GPEws}
\end{eqnarray}
with $c_{-2}=c_{2}=0$. The first term is simply the energy of site
$s$, the second term {[}with $U=g\int dxw^{4}(x)${]} accounts for
interactions of particles at the same site, and the two last terms
correspond to the exchange of particles between neighbor sites, with~%
\footnote{We work in the first-neighbors approximation which is valid if $V_{0}/F\gtrsim5$
. In this limit, we can neglect terms proportional to $\int dxw^{2}(x)w^{2}(x+1)\ll\int dxw^{3}(x)w(x+1)$.%
}
\begin{align}
J_{+}= & g\int dxw^{3}(x)w(x+1)\nonumber \\
J_{-}= & g\int dxw(x)w^{3}(x+1).\label{eq:J+J-}
\end{align}
 It is useful to write the wave function components as amplitude-phase
variables, defined by $c_{s}=\sqrt{I_{s}}e^{i\theta_{s}}$.

\begin{figure*}
\selectlanguage{french}%
\centering{}\includegraphics[width=5.2cm]{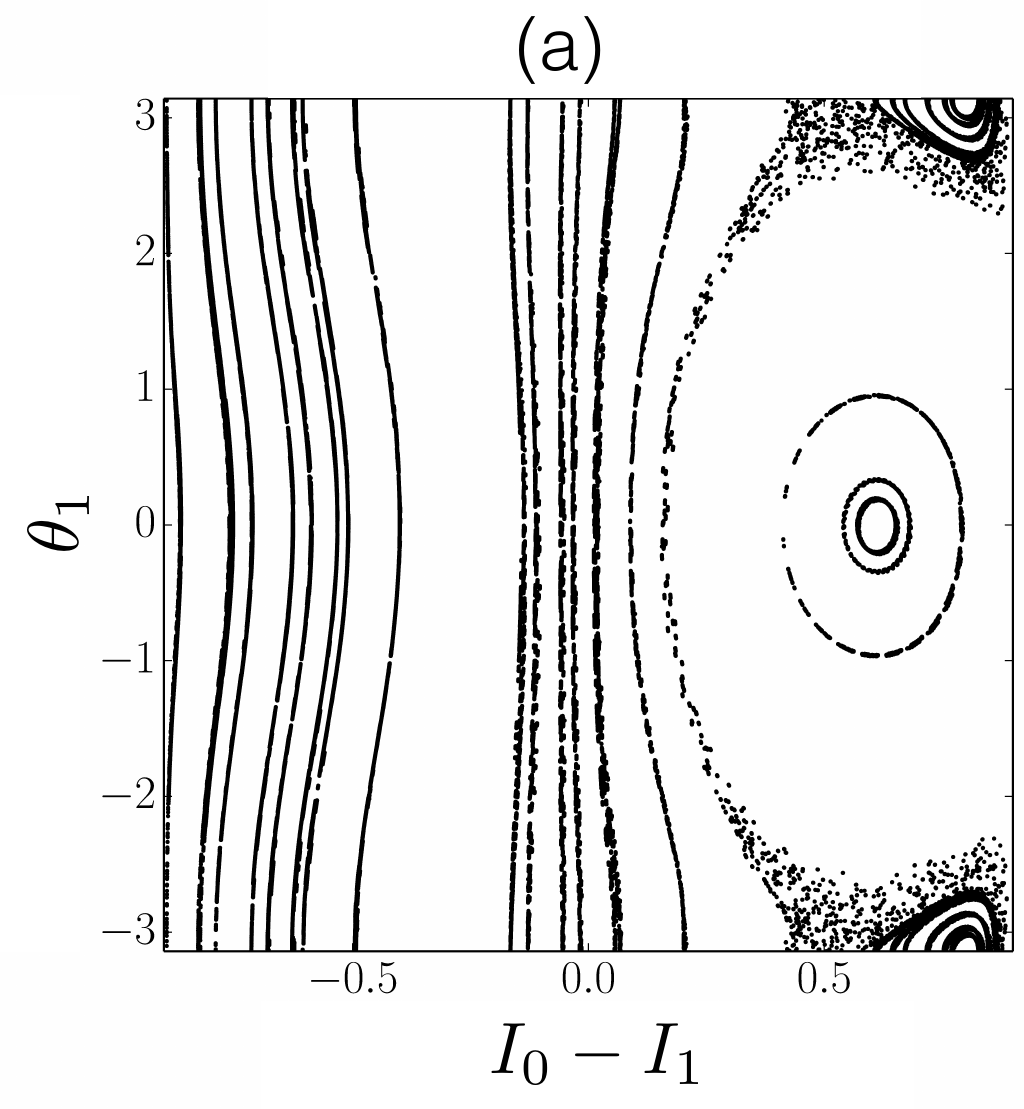}\qquad\includegraphics[width=5.5cm]{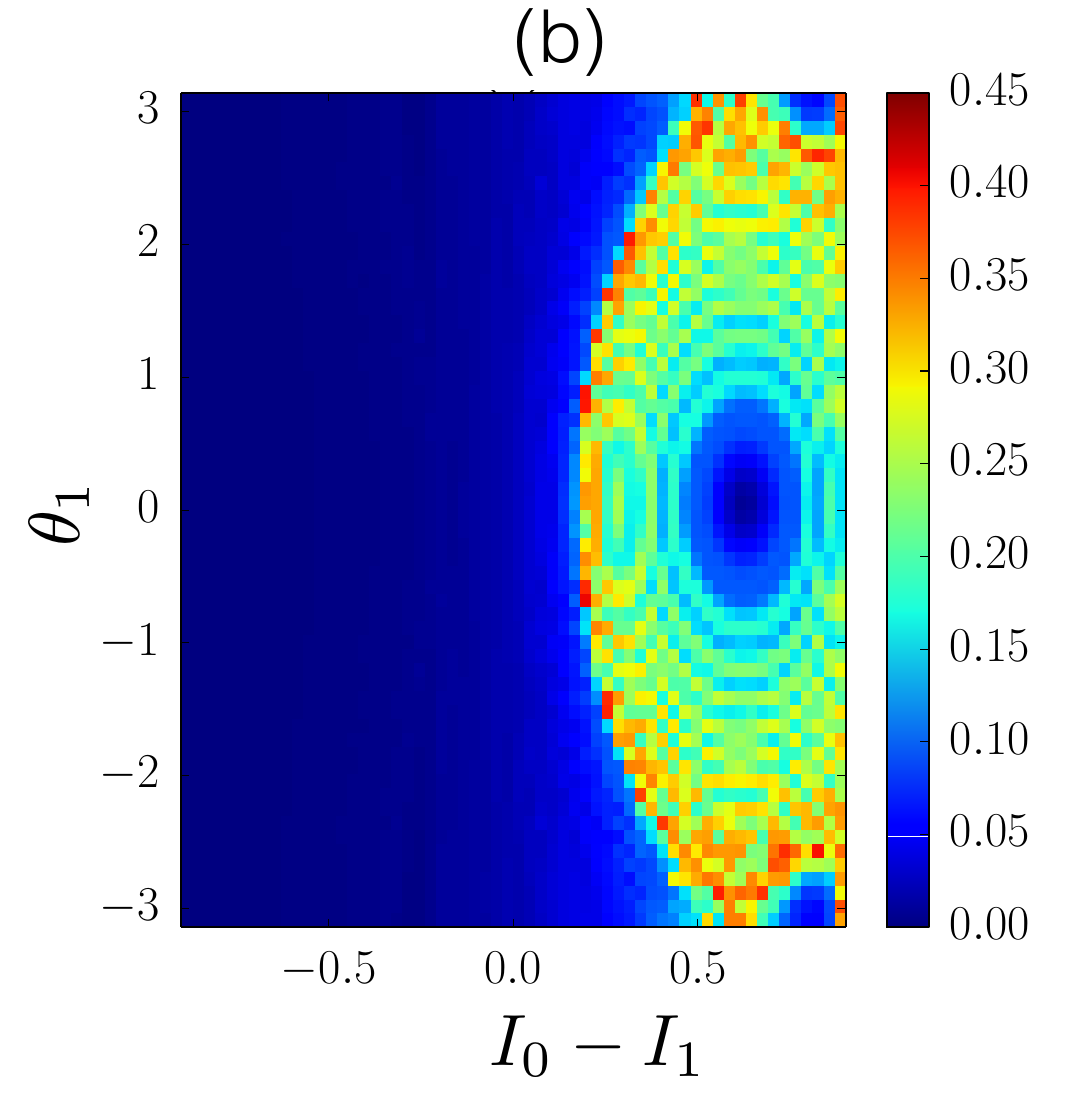}\foreignlanguage{english}{\protect\caption{\label{fig:Poincare}Quasiclassical chaos in a tilted lattice, obtained
by integrating the Gross-Pitaevskii equation~\eqref{eq:GPEws}. (a)
Poincar{\'e} section corresponding to $I_{-1}=0.1$, $\theta_{0}-\theta_{-1}=0$.
(b) Spectral entropy calculated between $t=0$ and $t=800$. Parameters
are $V_{0}=5$, $F=0.25$, $gN=0.2$.}
}\selectlanguage{english}%
\end{figure*}

The dynamics of such system has been studied in previous works~\cite{Thommen:ChaosBEC:PRL03,Lepers:QuasiClassTrack:PRL08,Kolovsky:BlochOscillationsBECDynamicalInst:Quantum:PRA09,Kolovsky:BECsOnTiltedLattices:PRA10},
where it was in particular showed that quasiclassical chaos in the
above system has the structure prescribed by the Kolmogorov-Arnold-Moser
theorem, which becomes apparent in a Poincar{\'e} section of the
dynamics, Fig.~\ref{fig:Poincare}a. We represented in Fig.~\ref{fig:Poincare}b
the same dynamics described in terms of the spectral entropy $S$.
Given a dynamical function $f(t\in[0,t_{\max}])$ (e.g. the average
position $\left\langle x\right\rangle (t)$), the spectral entropy
is related to the power spectrum $\tilde{F}(\nu)$ of $f(t)$ by
\[
S=-\frac{1}{\ln n_{\nu}}\sum_{\nu}\tilde{F}(\nu)\ln\tilde{F}(\nu),
\]
where $n_{\nu}=1+t_{\max}/\delta t$ is the number of frequency components
of $f(t)$ (with $\delta t$ the corresponding resolution). Roughly
speaking, the spectral entropy gives the number of significant frequency
components present in the spectrum, and is thus a good indicator of
(quasi)classical chaos. \textcolor{black}{The aim of the present work,
is to understand how the (linear) dynamics of the exact many-body
problem can approach the (quasiclassical) chaotic behavior observed
in Fig}\textcolor{blue}{.~\ref{fig:Poincare}.}

\section{Numerical approaches\label{sec:Numerical-approaches}}

The exact many-body problem can be described by the Bose-Hubbard Hamiltonian~\cite{Jaksch:ColdBosonicAtomsInOpticalLattices:PRL98,Cohen-TannoudjiDGO:AdvancesInAtomicPhysics::11,Meystre:AtomOptics:01}.
The single-particle Hamiltonian corresponds to Eq.~(\ref{eq:H1WS}),
whereas binary atom-atom contact interactions arise from a term $(g/2)\int dx\hat{\psi}^{\dagger}(x)\hat{\psi}^{\dagger}(x)\hat{\psi}(x)\hat{\psi}(x)$
where $\hat{\psi}(x)$ is the matter-wave field, which is expanded
in the Wannier-Stark basis $\hat{\psi}=\sum_{s}w(x-s)a_{s}$ where
$a_{s}$ is the boson annihilation operator for the site $s$. This
leads to a Bose-Hubbard Hamiltonian
\begin{equation}
\begin{array}{ccc}
H_{BH} & = & \sum_{s}\left[Fsn_{s}+\frac{U}{2}n_{s}(n_{s}-1)\right.\\
 &  & \left.+\left(J_{+}a_{s}^{\dagger}a_{s+1}n_{s}+J_{-}a_{s}^{\dagger}a_{s-1}n_{s}+\mathrm{h.c.}\right)\right]
\end{array}\label{eq:HBH}
\end{equation}
where $n_{s}=a_{s}^{\dagger}a_{s}$ is the number operator for site
$s$. Note that the two ``kinetic energy'' terms on the second line
are different from those of the usual Bose-Hubbard Hamiltonian as,
contrary to Wannier functions of a periodic potential, Wannier-Stark
functions are \emph{eigenstates} of the one-particle Hamiltonian Eq.~\eqref{eq:H1WS}.
Transport here is due to \emph{interactions} between neighbor sites,
not to tunnel effect, and is thus proportional to the interaction
strength $g$ {[}cf. Eq.~(\ref{eq:J+J-}){]}. One can convince oneself
that the Bose-Hubbard Hamiltonian describes the many-body Wannier-Stark
problem by noticing that the mean-field approximation which consists
in replacing the operators $a_{s}$ by c-numbers $c_{s}$ leads to
a classical Hamiltonian whose equations of motions correspond to the
GPE,~Eq.(\ref{eq:GPEws}). In the following, we will compare the
solution of the many-body Schr{\"o}dinger equation
\begin{equation}
i\hbar\frac{\partial|\psi_{BH}\rangle}{\partial t}=H_{BH}|\psi_{BH}\rangle,\label{eq:phiBH}
\end{equation}
to the solution of Eq.~(\ref{eq:GPEws}). It is challenging to directly
diagonalize Eq.~(\ref{eq:HBH}) which is of dimension $(N+1)(N+2)/2\sim N^{2}$
for atom numbers $N$ larger than a few tens, so we used the so-called
``Lanczos diagonalization'' (LD)~\cite{Lanczos:MethodForTheSolutionEigProblem:JRNBS14,Manmana:TimeEvolutionQuantumManyBody:ACP05,Carleo:LocalizationManyBodySystems:SR14},
which consists in using a truncated many-body evolution operator over
a short time interval $dt$:
\begin{equation}
U(dt)=\sum_{n=0}^{n_{K}}\frac{\left(-iH_{BH}dt\right)^{n}}{n!}.\label{eq:truncUt}
\end{equation}
In the Lanczos scheme, in order to evolve the wave function $\left|\psi_{BH}(t)\right\rangle $
during $dt$ one first builds an orthonormal basis $|l_{1}\rangle,..,|l_{n_{K}}\rangle$
of the subspace spanned by $n_{K}$ ``Krylov vectors'' defined as
$|\psi_{BH}(t)\rangle,H|\psi_{BH}(t)\rangle,H^{2}|\psi_{BH}(t)\rangle,..,H^{n_{K}}|\psi_{BH}(t)\rangle$.
The resulting tridiagonal Hamiltonian of dimension $n_{K}\sim10-20\ll N^{2}$
can be then diagonalized to obtain the Lanczos eigenvectors $|\nu_{1}\rangle,..,|\nu_{n_{K}}\rangle$
and eigenvalues $\alpha_{1},..,\alpha_{n_{K}}$. The propagation of
the wave function is then approximated as:
\[
|\psi_{BH}(t+dt)\rangle=\sum_{i=1}^{n_{K}}\langle\nu_{i}|\psi_{BH}(t)\rangle e^{-i\alpha_{i}dt}|\nu_{i}\rangle.
\]
As the error in this approximation can be controlled by adjusting
the time step $dt$, the Lanczos procedure is considered as an ``exact
diagonalization'' technique. Each time step implies the diagonalization
of a $n_{K}\times n_{K}$ matrix thus the method is only interesting
if $n_{K}\ll N^{2}$\textcolor{magenta}{. }The Lanczos procedure allows
us to treat the many-body problem with atom numbers $N\lesssim500$
in reasonable computer times.

Describing the system by a set of complex numbers $c_{s}$ instead
of quantum operators $a_{s}$, the GPE potentially neglects two important
effects: (i) quantum noise, i.e the quantum fluctuations around the
expectation value of $a_{s}$ (ii) quantum interferences, the fact
that two different trajectories can interfere %
\footnote{We use here the language of second quantization which means that quantum
interference is described by GPE only at the single particle level
(e.g. it displays nonlinear Bloch Oscillations) - that is, interference
governed by the phase of the $c_{s}$ themselves. %
}. The truncated Husimi method (THM) allows one to reinsert quantum
noise in a GPE description by propagating independently sets of\textcolor{red}{{}
}initial conditions whose distribution reflects the quantum noise.
However, the THM neglects quantum interference effects that may occur
between the different trajectories and is thus an intermediary approach
between the GPE and the exact solution of the many-body system particularly
suited to identify the role of the quantum noise~\cite{Sinatra:TruncatedWignerBEC:JPBAMOP02,Trimborn:MeanFieldlBoseHubbard:PRA09,Trimborn:ExactNumberConservingBoseHubbard:PRA08}.
One thus needs to choose a rule relating the complex variables $\{c_{s}^{\prime}\}$
serving as initials conditions to the GPE to the many-body problem,
i.e to choose a family of many-body states able to connect a many-body
state $|\Omega\left(\left\{ c_{s}\right\} \right)\rangle$ and the
corresponding set of GPE initial conditions $\{c_{s}^{\prime}\}$.
With $M$ sites ($M=3$ in this work), one can choose the so-called
$SU(M)$ coherent states~\cite{Zhang:CoherentStates:TheoryApplications:RMP90}
with finite total number of atoms $N$, which are defined, in the
Fock basis, as
\[
|\Omega\left(\left\{ c_{s}\right\} \right)\rangle=\sqrt{N!}\sum_{n_{1}+..+n_{M}=N}\prod_{s}\frac{c_{s}^{n_{s}}}{\sqrt{n_{s}!}}|n_{1},..,n_{s}\rangle
\]
with $\sum_{s}|c_{s}|^{2}=1$. The advantage of $SU(M)$ states, as
compared to the usual Glauber coherent states, is that the \emph{total}
number of atoms $N$ is fixed: $\langle\left(\hat{N}-\left\langle \hat{N}\right\rangle \right)^{2}\rangle=0$,
with $\hat{N}=\sum_{s}n_{s}$, which is a situation closer to the
that encountered in a real experiment. We then need to choose a quantity
to characterize the quantum noise in the system. In this work we use
the Husimi function $Q_{\Omega}$ which, given a coherent state $|\Omega\left(\left\{ c_{s}\right\} \right)\rangle$,
is simply defined as its projection over all possible states: $Q_{\Omega}\left(\left\{ c_{s}^{\prime}\right\} \right)=|\left\langle \Omega^{\prime}\left(\left\{ c_{s}^{\prime}\right\} \right)\right|\left.\Omega\left(\left\{ c_{s}\right\} \right)\right\rangle |^{2}$.
The width of this function decreases as $N^{-1/2}$ and thus gives
a good representation of the quantum noise. For a given initial many-body
state $|\Omega\left\{ c_{s}\right\} \rangle$, we generate the corresponding
\emph{distribution} of one-particle initial conditions $D\left(c_{s}\right)\equiv\{c_{s}^{\prime}\}$
which mimics the Husimi representation of the $|\Omega\left\{ c_{s}\right\} \rangle$.
Each of these states is then propagated independently according to
the GPE and observables are obtained by averaging over the distribution
of final states. For large enough $N$, the width of the Husimi distribution
tends to $0$ and one recovers the results of the GPE. Hence, the
THM allows one to use the GPE to simulate the evolution of not so
large number of atoms (including quantum noise but not interference
effects), and it turns out to be a convenient tool to study of the
emergence of chaos in our system as the number of atoms increases.

\section{Emergence of quasiclassical chaos}

\begin{figure}
\begin{centering}
(a)\includegraphics[width=0.95\columnwidth]{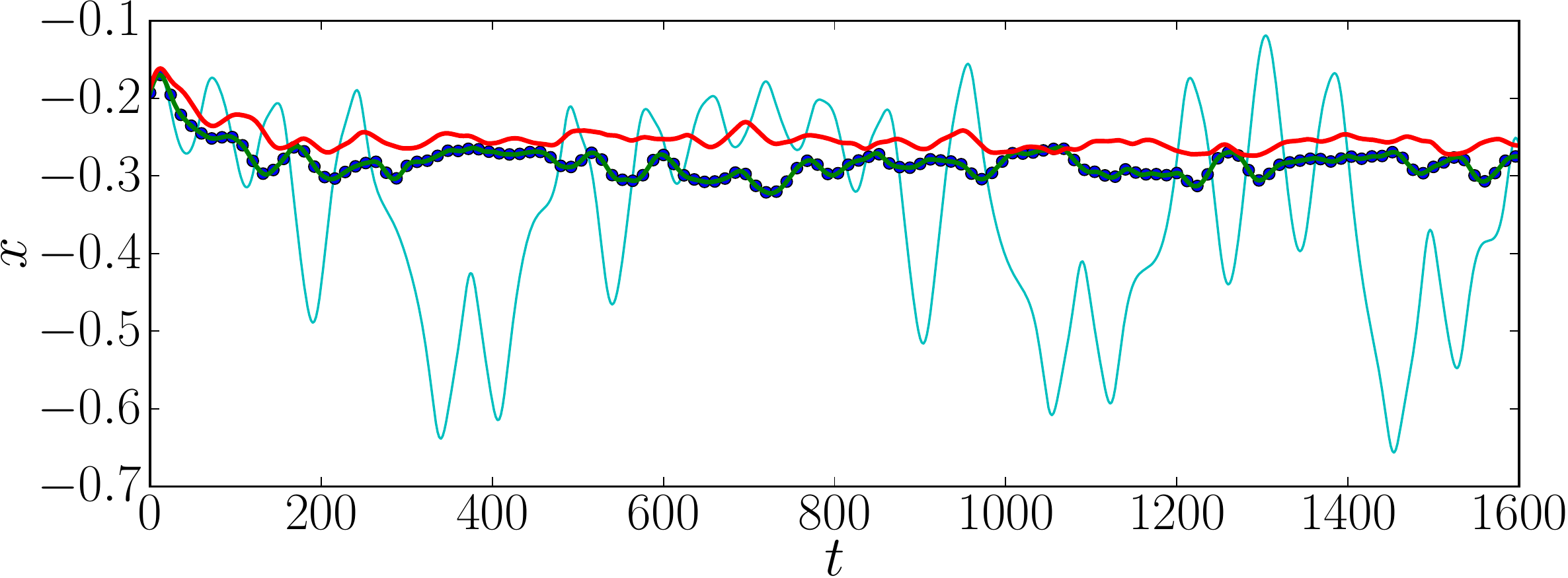}
\par\end{centering}

\centering{}(b)\includegraphics[width=0.95\columnwidth]{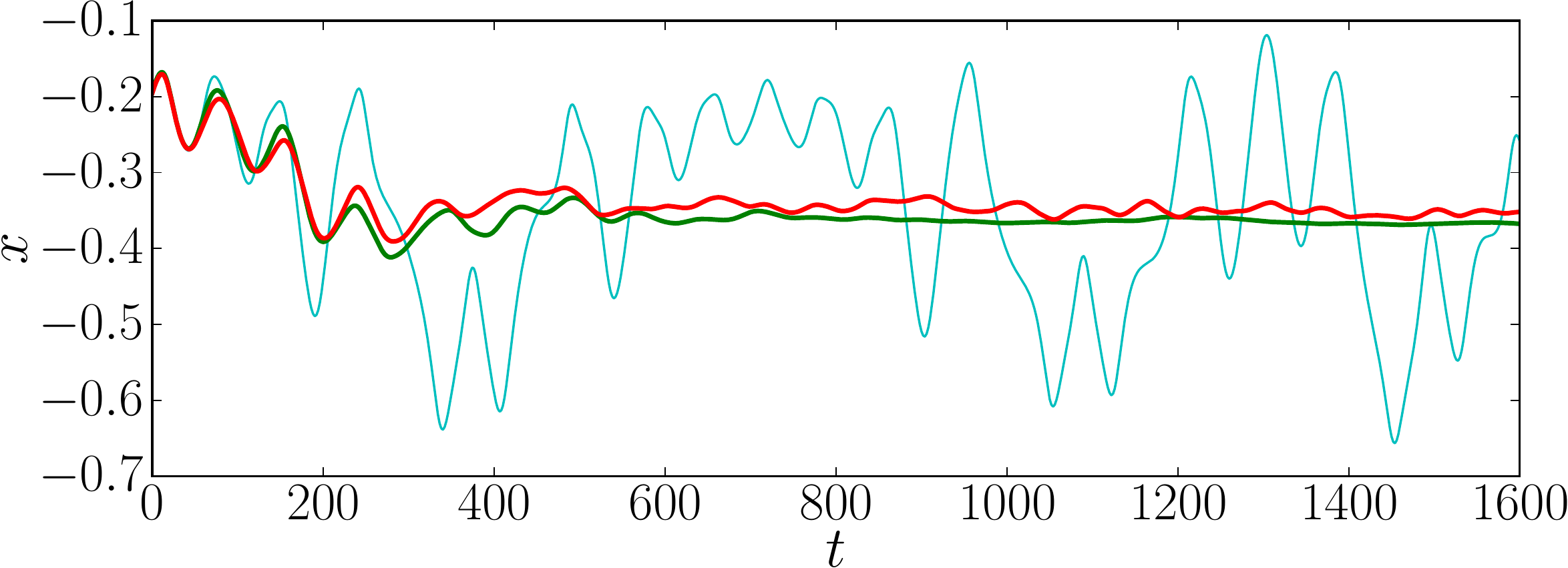}\protect\caption{\label{fig:AveragePosition}Evolution of the average position of a
wave packet calculated with different methods: Gross-Pitaevskii equation
(cyan), Lanczos diagonalization (green), and the truncated Husimi
method (red), for (a) $N=30$ and (b) $N=500$. In panel (a), we also
represented by blue circles the result obtained by direct diagonalization
of the full many-body problem, which is in perfect agreement with
the result of the Lanczos method, illustrating the accuracy of the
later. The initial state is a $SU(3)$ coherent state with $c_{-1}=\sqrt{0.5},$
$c_{0}=\sqrt{0.25}$ and $c_{1}=i\sqrt{0.25}$. Other parameters are
$gN=0.2$, $V_{0}=7$, $F=0.1$.}
\end{figure}

We compare in Fig.~\ref{fig:AveragePosition} the time evolution
of the sys\textcolor{black}{tem -- represented here by the average
position of the wave packet -- calculated accordin}g to the different
methods described in sec.~\ref{sec:Numerical-approaches}, for a
small $N=30$ {[}plot (a){]} and a larger (but not large) $N=500$
{[}plot (b){]} number of atoms. In order to have a pertinent comparison
with the GPE, $gN$ is the same for all simulations. The GPE calculation
displays a clear irregular behavior associated to the presence of
quasiclassical chaos. No such behavior is observed for $N=30$ atoms
either with the LD or the THM\textcolor{blue}{. }In plot (b), with
$N=500$, one sees that the GPE result ``sticks'' to the many-body
calculations for times $t\lesssim150$, and that the many-body calculations
even display the first oscillations observed in the GPE evolution.

\begin{figure}
\begin{centering}
\includegraphics[width=0.45\columnwidth]{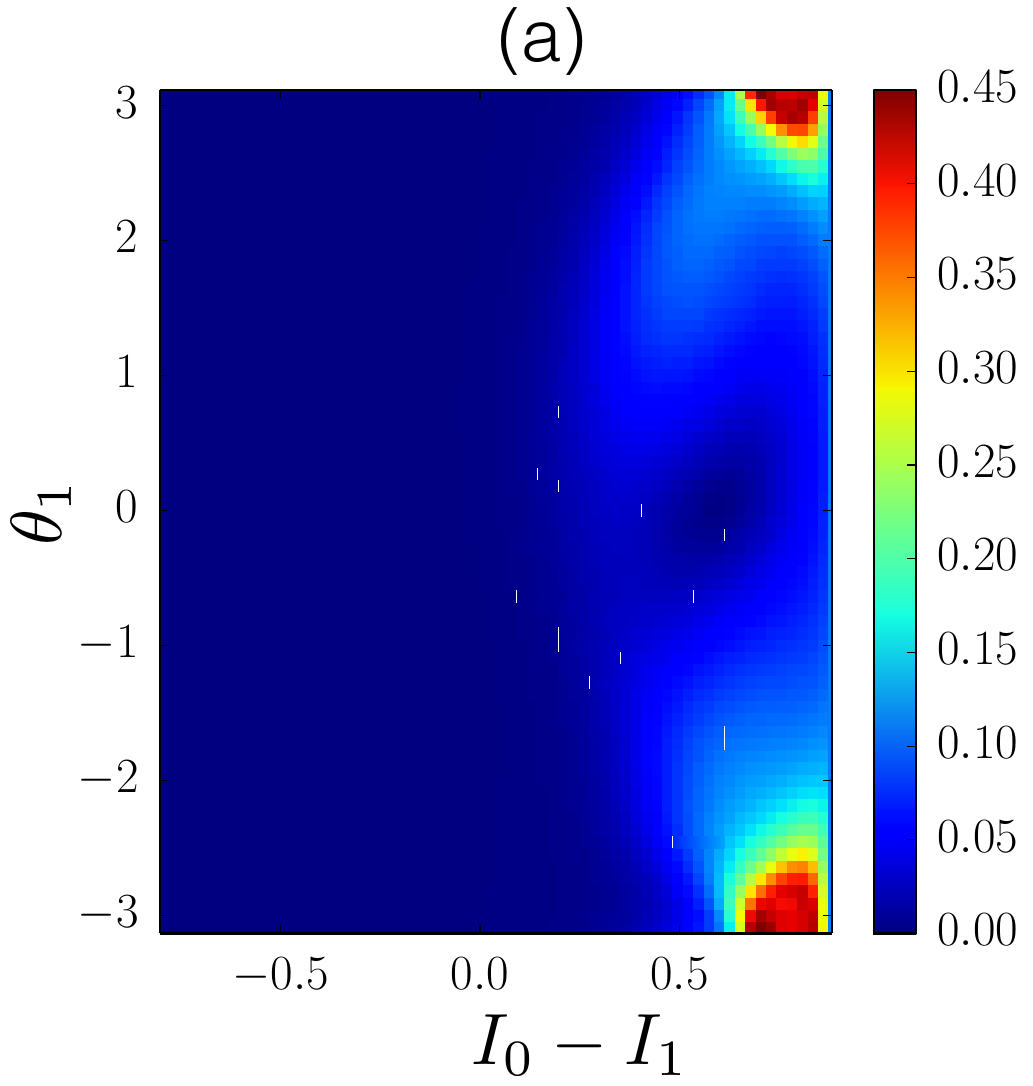}\qquad\includegraphics[width=0.45\columnwidth]{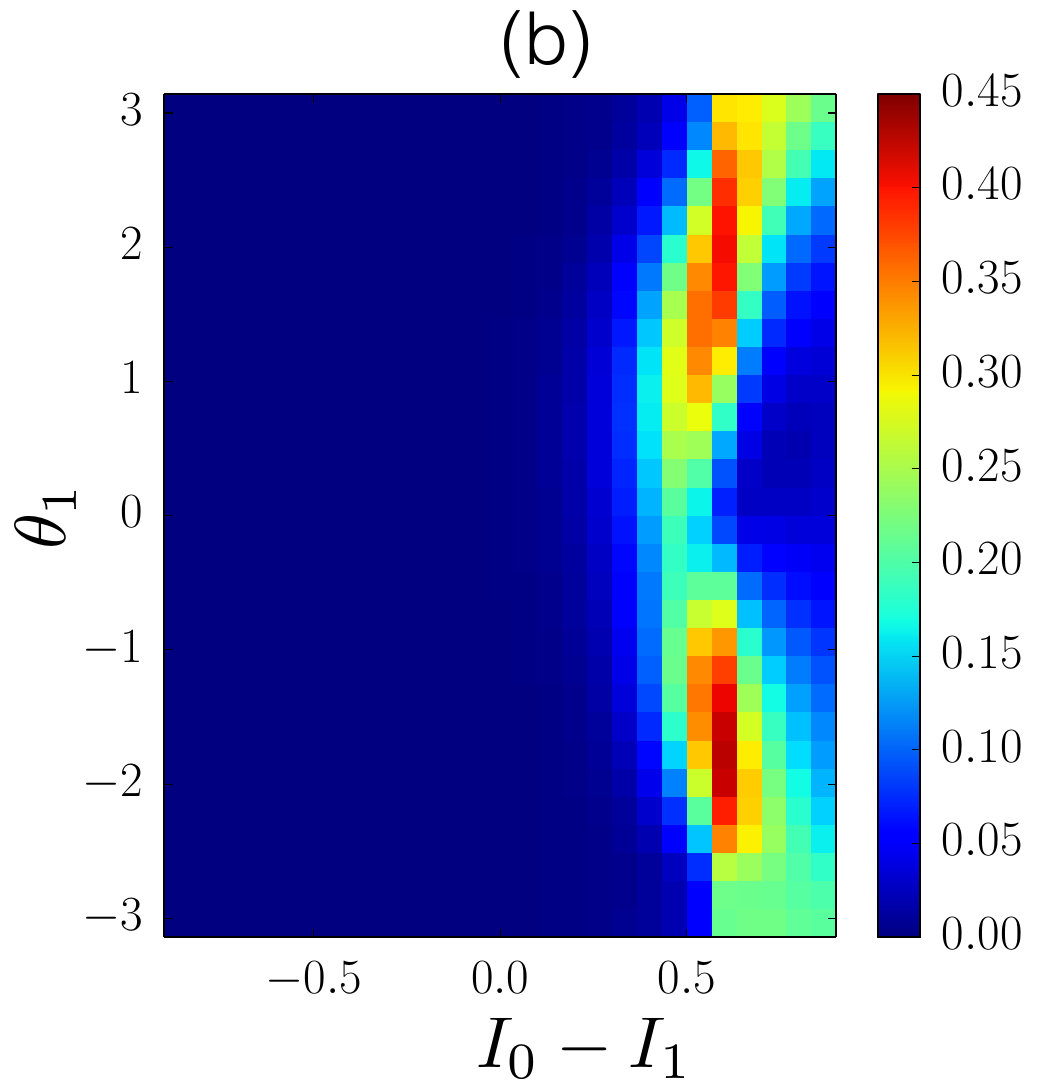}\vspace{0.5cm}

\par\end{centering}

\begin{centering}
\includegraphics[width=0.45\columnwidth]{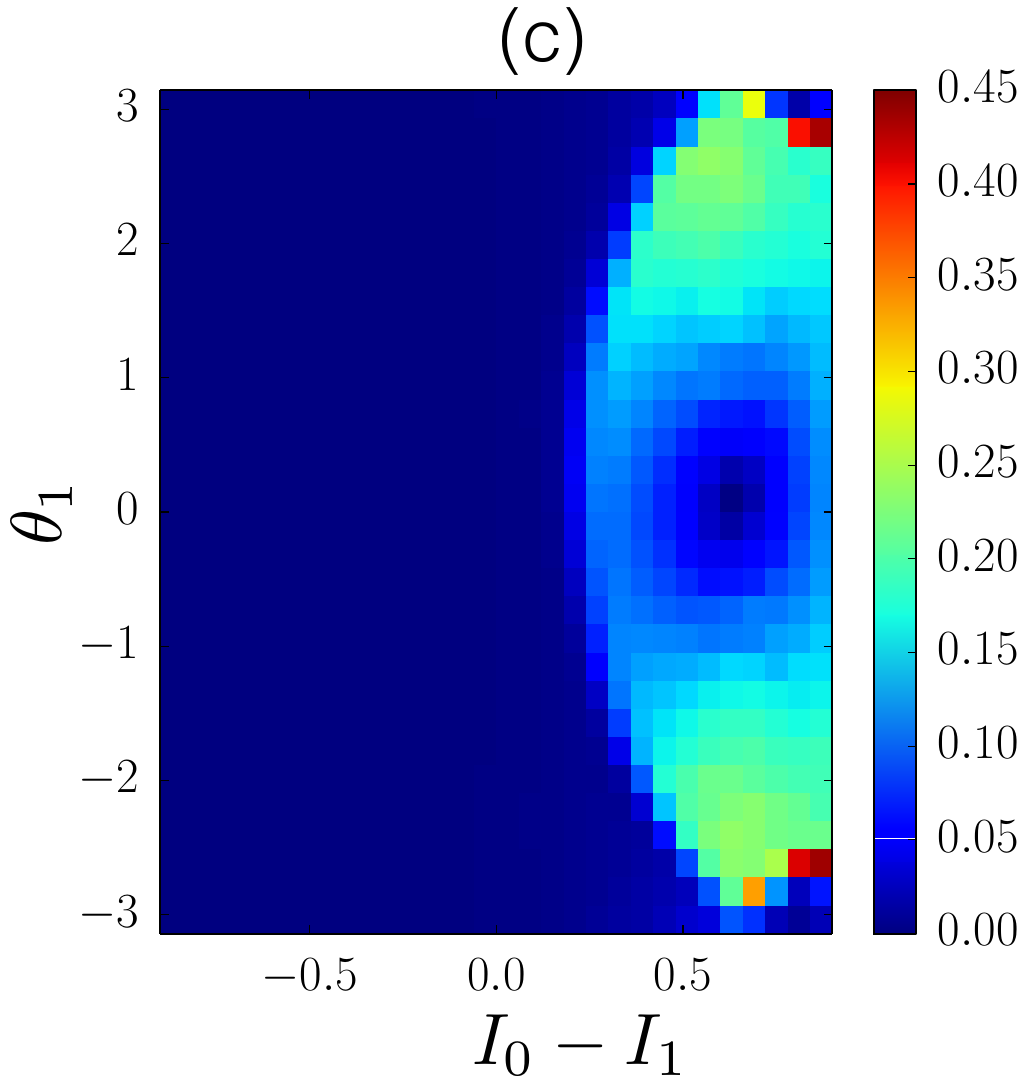}\qquad\includegraphics[width=0.45\columnwidth]{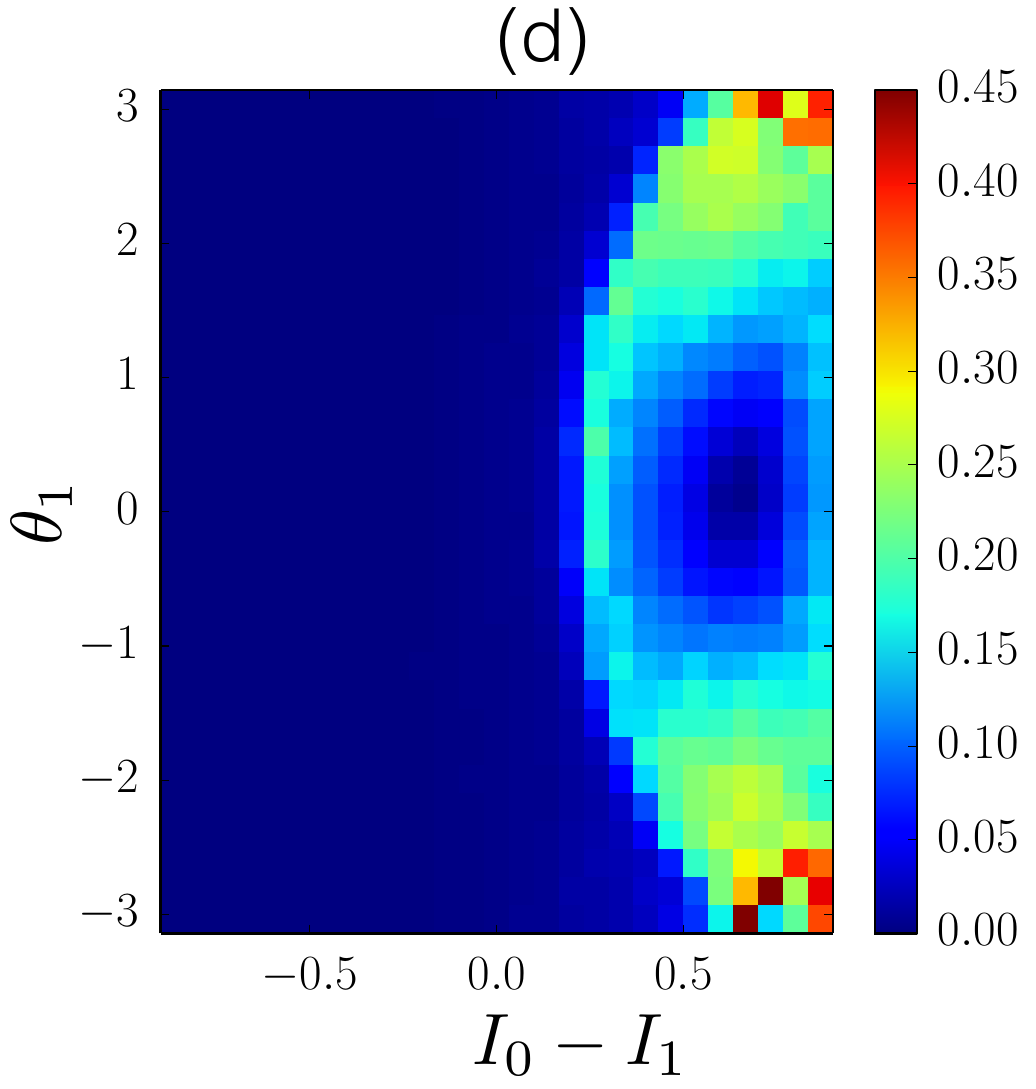}
\par\end{centering}

\protect\caption{\label{fig:SpectralEntroyDist}Spectral entropy distribution in phase
space obtained by LD (left column) and by THM (right column) with
$N=30$ (top row) $N=400$ (bottom row) and same parameters as Fig.\textcolor{blue}{~\ref{fig:Poincare}}.
When $N$ increases, the THM provides a quite good representation
of the mean-field behavior for large $N$, and the phase-space structure
of the many-body problem becomes similar to the Poincaré section obtained
from the GPE, plot (a) of Fig.~\ref{fig:Poincare}.}
\end{figure}

In order to observe the emergence of quasiclassical chaos as the number
of particles increases, we calculated the spectral entropy for a complete
set of initial $SU(3)$ states using both LD and THM. The width of
the Husimi function corresponding to a given initial condition is
determined, and we assume that for any initial condition $\left\{ c_{s}\right\} $,
the Husimi function can be approximated by a square distribution with
same width centered around $\left\{ c_{s}\right\} $. We checked numerically
that this approximation has no impact on our results. Figure~\ref{fig:SpectralEntroyDist}
compares the distribution of the spectral entropy computed by LD and
by the THM for $N=30$ and $N=400$ atoms, and shows that not only
the results become more similar when the atom number increases, but
also that the characteristic features observed in the quasiclassical
Poincaré section Fig.~\ref{fig:Poincare}a tend to emerge (this trend
is confirmed for intermediate atom numbers). These results show that
THM provides a good representation of the exact many-body behavior
for large atom numbers and can advantageously used for comparisons
with the quasiclassical behavior.

\begin{figure*}
\selectlanguage{french}%
\begin{centering}
\includegraphics[width=5cm]{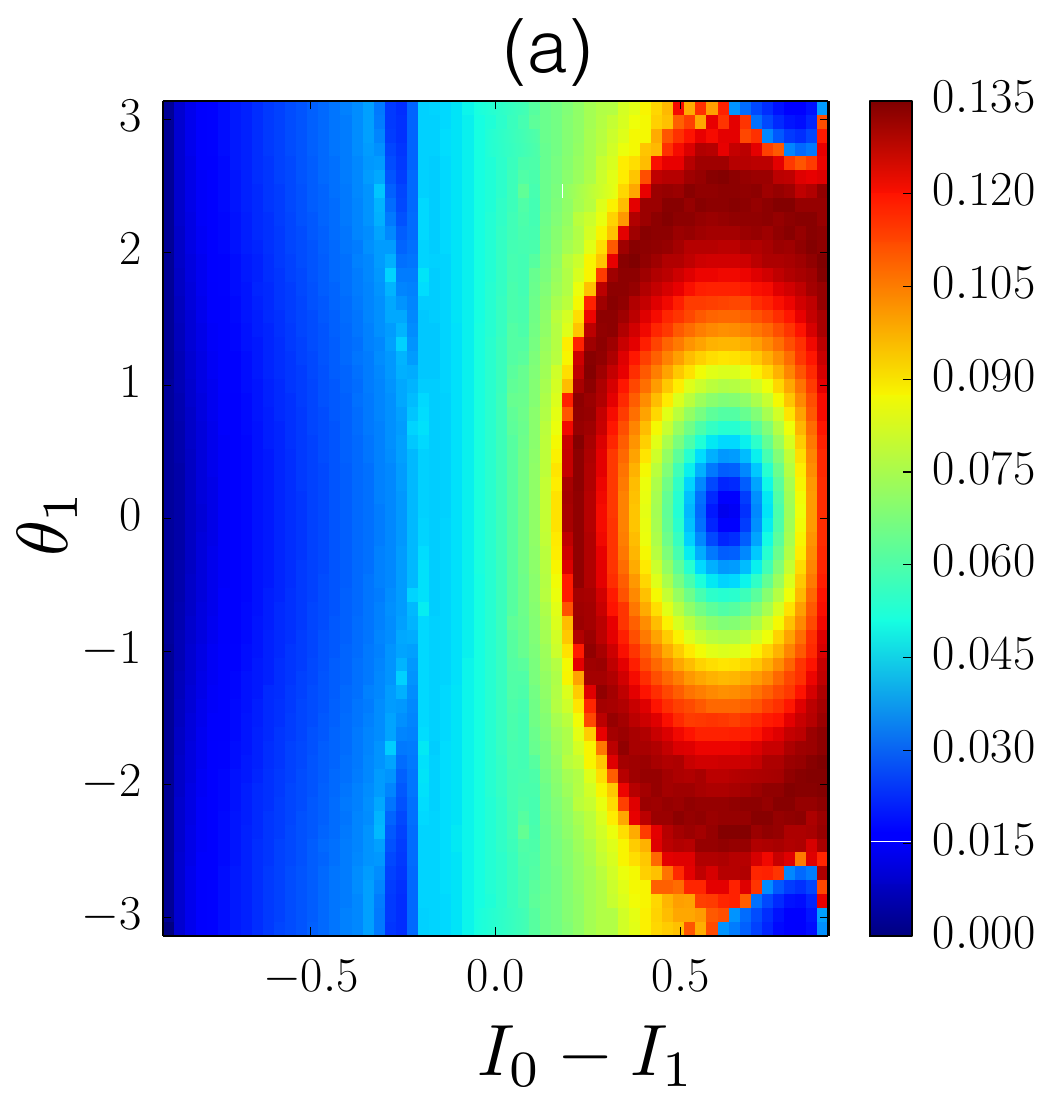}\quad\includegraphics[width=5cm]{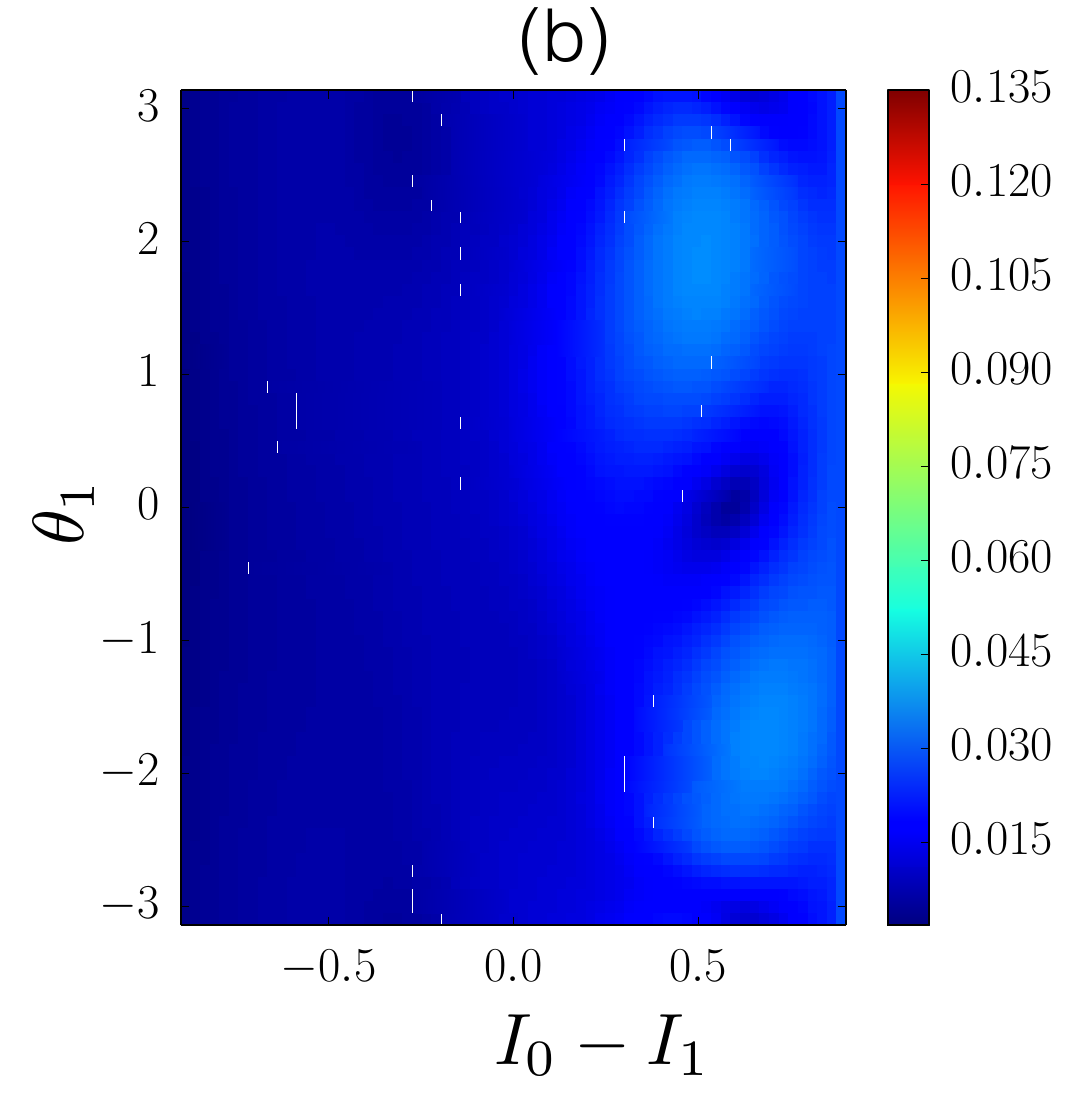}\quad\includegraphics[width=5cm]{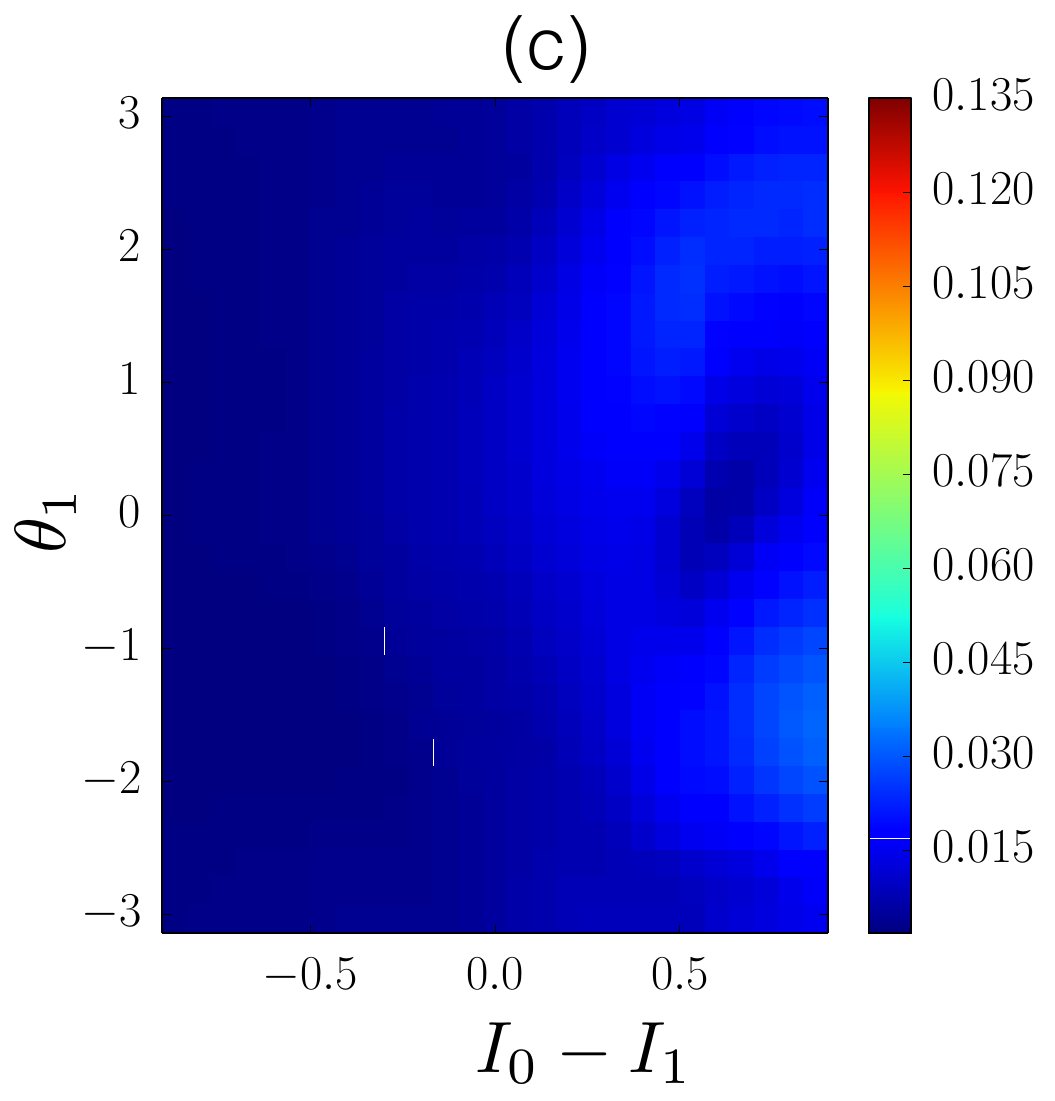}\vspace{0.5cm}
\par\end{centering}

\begin{centering}
\includegraphics[width=5cm]{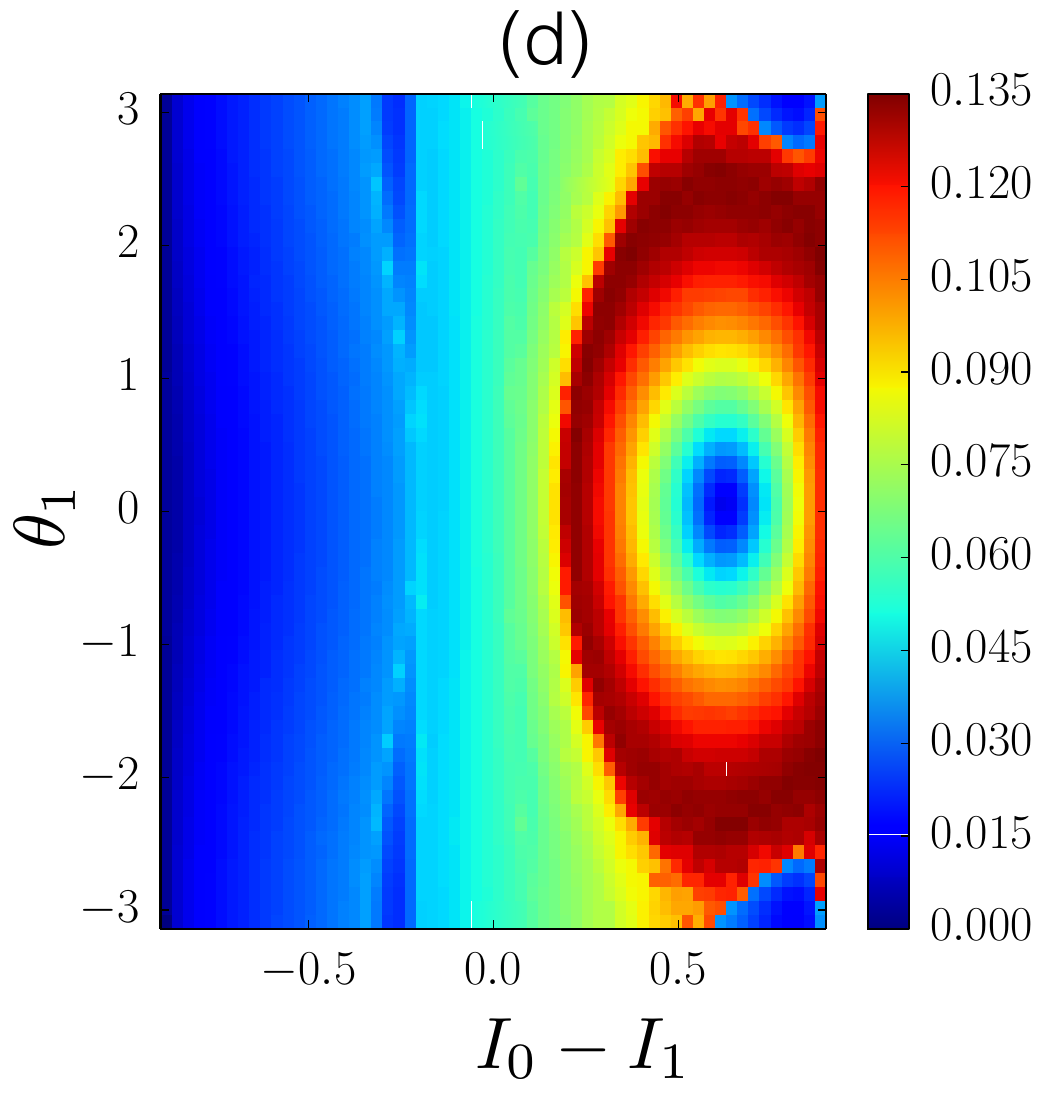}\quad\includegraphics[width=5cm]{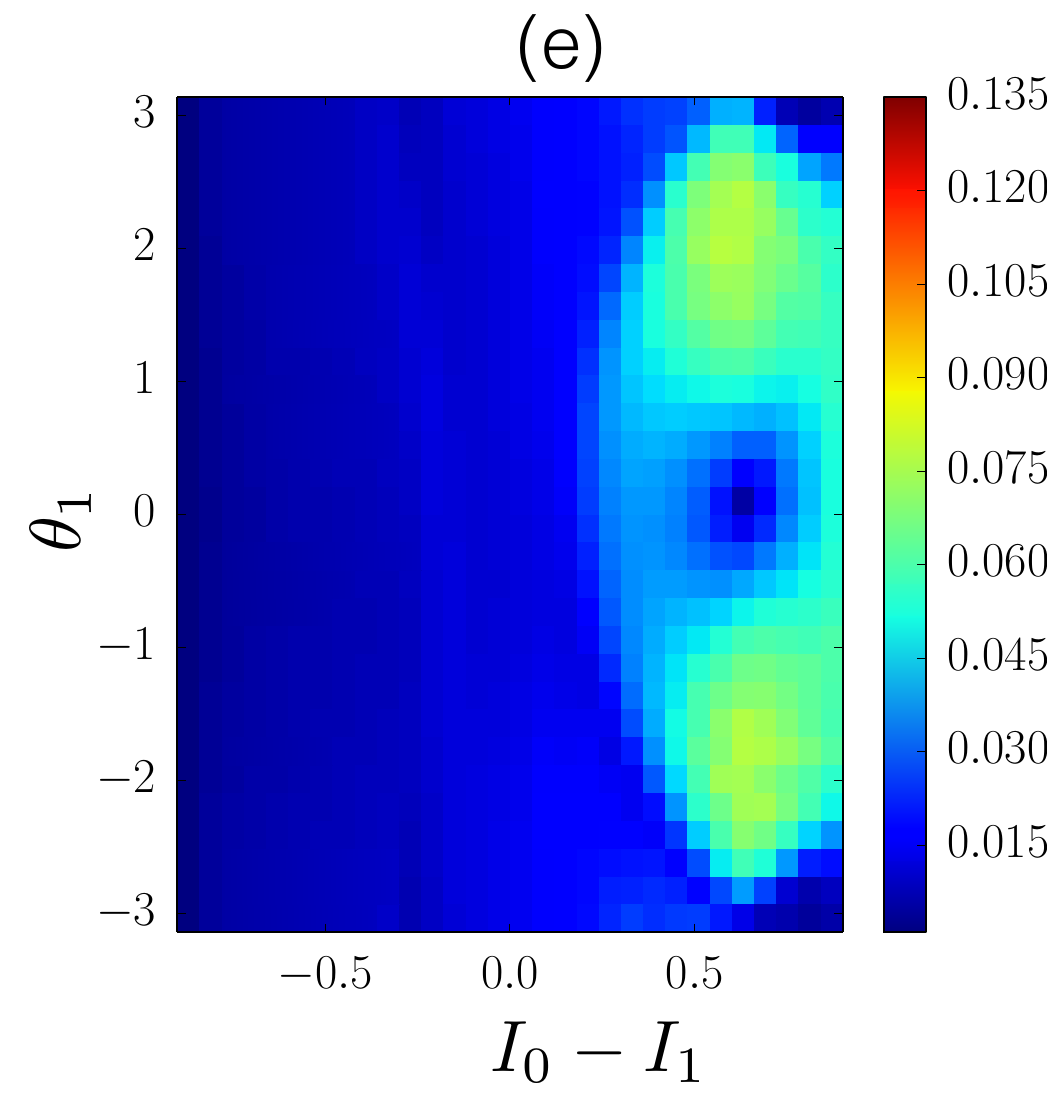}\quad\includegraphics[width=5cm]{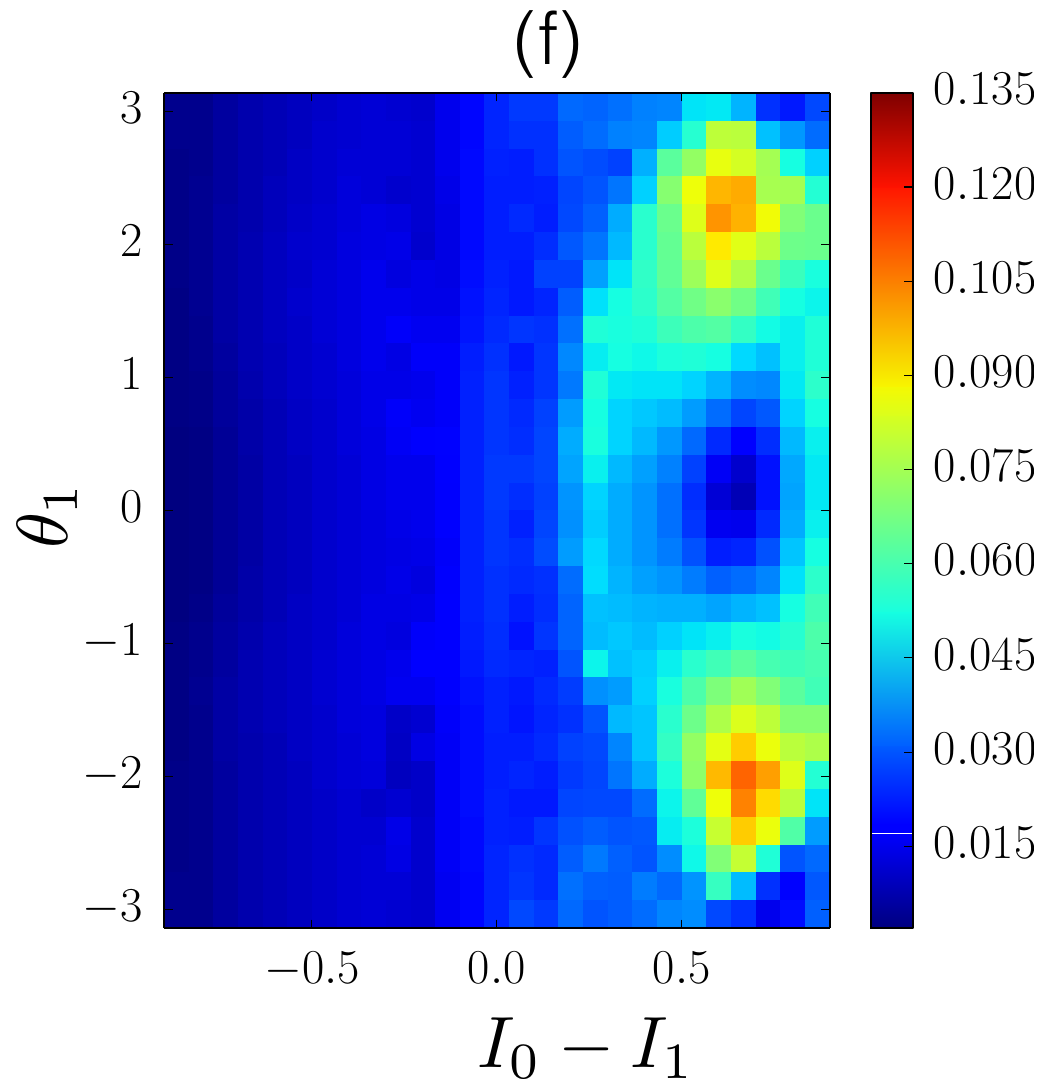}
\par\end{centering}

\selectlanguage{english}%
\protect\caption{\label{fig:StandardDeviationX} Variance of the wave packet position
calculated between $t=0$ and $t=800$ from the GPE (left column),
LD (center column), and THM (right column) with $N=30$ (top row)
and $N=400$ (bottom row) and the same parameters as Fig\textcolor{blue}{.~\ref{fig:Poincare}}.}
\end{figure*}

In figure~\ref{fig:StandardDeviationX} we present a calculation
of the variance of the wave packet position -- which is an indication
of the erratic character of the dynamics, that is quasiclassical chaos
-- by the three methods and for $N=30$ (top row) and $N=400$ (bottom
row). The GPE picture (first column) is obviously independent of the
atom number (provided $gN$ is constant). Both LD (center column)
and THM (right column) instead show a clear dependence in $N$ and
are rather different, both in shape as in the amplitude, from the
GPE result, which shows that quasiclassical chaos is not fully developed
for such atom numbers. However, one can see however that the higher
the number of atoms the closer the result is from the one obtained
with GPE. The amplitude of the variances increases by a factor 2 for
THM and by a factor 3.8 for the LD when the number of atoms is changed
from 30 to 400, and the surface of the chaotic zones (in yellow and
red) also increases significantly. Thus, even if the atom numbers
considered here are clearly too small to allow a definite conclusion,
one can reasonably expect a much better convergence with the GPE for
larger values of $N$.

\section{Conclusion}

In conclusion, we showed that the dynamics exhibited by the many-body
problem as described by both exact diagonalization and the truncated
Husimi method tends to converge to the quasiclassical chaos displayed
by the mean-field approximation as the atom number increases. Our
result suggests the existence of a ``nonlinear characteristic time''
increasing with some (monotonous) function of $N$, during which the
prediction of the mean-field approximation and of the exact many-body
problem agree. One may conjecture that this time is related to the
typical time for particle number fluctuations to affect significantly
the dynamics, but we are presently unable to give a verifiable estimation
of such time. In particular, THM appears to approach very well the
exact dynamics, which suggests that the essential effect leading to
the emergence of quasiclassical chaos as the number of particles increases
is the reduction of quantum noise. It would be interesting to analyze
the influence of quantum interferences and diffusion as done e.g.
by Carvalho \emph{et al}.~\cite{Carvalho:DecoherncePhaseSpaceKHO:PRE04},
who compared the phase space structure of the classical chaotic \emph{one-body
}delta-kicked harmonic oscillator to the Wigner representation of
its quantum counterpart, and showed that both decoherence and classical
diffusion lead to a similarity of the two representations. Weiss and
Teichmann~\cite{Weiss:DifferencesMeanFieldQuantumDynamics:PRL08},
using a \emph{many-body} system (with $N=1000$) slightly different
from ours, attributed the observed differences between the mean-field
and the many-body dynamics to the existence of entanglement in the
many-body problem, and showed that these differences are reduced if
decoherence is added to the system. The present work sheds a different
light on the problem of the quantum-classical transition in a system
displaying quasiclassical chaos, suggesting the quantum noise is the
main cause of the differences. In order to reach more definitive conclusions,
however, simulations of the many-body problem with atom numbers larger
by (at least) one order of magnitude would be necessary. This is a
fascinating yet challenging goal for future work.
\begin{acknowledgments}
Laboratoire de Physique des Lasers, Atomes et Molécules is UMR 8523
of CNRS. Work partially supported by Agence Nationale de la Recherche
(grants LAKRIDI ANR-11-BS04-0003,2 K-BEC ANR-13-BS04-0001-01) and
the Labex CEMPI (ANR-11-LABX-0007-01). JCG thanks the Max-Planck Institute
for the Physics of Complex Systems, Dresden, for support in the framework
of the Advanced Study Group on Optical Rare Events.\bibliographystyle{apsrev}
\bibliography{ArtDataBase}
\end{acknowledgments}

\end{document}